\documentclass[runningheads]{llncs}
\usepackage[sort]{cite}
\usepackage[T1]{fontenc}
\usepackage{graphicx}
\usepackage{booktabs}
\usepackage{enumitem}
\usepackage{subcaption}
\usepackage[autostyle]{csquotes}

\usepackage{xcolor}
\usepackage{soul}
\usepackage[square,numbers]{natbib}
\begin{document}

\newcommand{\add}[1]{#1}
\newcommand{\cut}[1]{\textcolor{pink}{#1}}
\newcommand{\del}[1]{}
\newcommand{\moved}[1]{\textcolor{black}{{#1}}}
\newcommand{\stats}[1]{\textcolor{black}{#1}}
\newcommand{\todo}[1]{\textcolor{red}{[todo: #1]}}

\newcommand{\myquote}[1]{\enquote{\textit{#1}}}

%
\title{Learner and Instructor Needs in AI-Supported Programming Learning Tools: Design Implications for Features and Adaptive Control}

\titlerunning{Design Implications for AI Programming Tools}
%
%

\author{Zihan Wu \and
Yicheng Tang\and
Barbara J. Ericson}
\authorrunning{Wu et al.}
%
\institute{University of Michigan
\email{\{ziwu,tangyc,barbarer\}@umich.edu}}
\maketitle              
\vspace{-12pt}
\begin{abstract}
AI-supported tools can help learners overcome challenges in programming education by providing adaptive assistance. 
However, existing research often focuses on individual tools rather than deriving broader design recommendations. 
A key challenge in designing these systems is balancing learner control with system-driven guidance.
To explore user preferences for AI-supported programming learning tools, we conducted a participatory design study with 15 undergraduate novice programmers and 10 instructors to gather insights on their desired help features and control preferences, as well as a follow-up survey with 172 introductory programming students.

Our qualitative findings show that learners prefer help that is encouraging, incorporates visual aids, and includes peer-related insights, whereas instructors prioritize scaffolding that reflects learners' progress and reinforces best practices. 
Both groups favor shared control, though learners generally prefer more autonomy, while instructors lean toward greater system guidance to prevent cognitive overload. 
Additionally, our interviews revealed individual differences in control preferences. 

Based on our findings, we propose design guidelines for AI-supported programming tools, particularly regarding user-centered help features and adaptive control mechanisms. 
Our work contributes to the human-centered design of AI-supported learning environments by informing the development of systems that effectively balance autonomy and guidance, enhancing AI-supported educational tools for programming and beyond.


\vspace{-10pt}

\end{abstract}

\section{Introduction}
\vspace{-6pt}

Programming is an important skill, but learning to program is not easy~\cite{milne2002difficulties}. 
To address this problem, researchers have designed various tools to provide programming help such as intelligent tutors~\cite{crow2018intelligent}, programming games \cite{lindberg2019gamifying}, and practice environments with adaptive hints and explanations~\cite{parsons2006parson, ericson2023multi}.
AI-supported tools, in particular, have shown promise in offering personalized guidance and adaptive feedback~\cite{le2013review, kazemitabaar2024codeaid, taneja2024jill, caughey2023investigating}.
However, most prior research has focused on designing and evaluating individual tools rather than exploring broader design guidelines regarding the needs and expectations of novice learners and instructors.
In this work, we focus our research on AI-supported learning tools that provide on-demand help when learners are writing code.

To assist future work in better designing learner-centered AI-supported programming learning tools, we investigate the desired \textbf{help features}, referring to the abstract qualities of help that users find beneficial.
Our first research question asks: \textbf{RQ1: What \textit{help features} do novice learners and instructors desire from an adaptive learning tool that provides on-demand programming help?}

A key challenge in AI-supported adaptive learning systems is determining
how much decision-making power and responsibility should be given to the learner versus the system~\cite{imhof2020implementation}.
If the system has full control, learners might feel frustrated when the system's intervention does not align with their expectations; if the learners have full control, they might be overwhelmed~\cite{imhof2020implementation, nwana1990intelligent}.
While some prior work has investigated the impact of control on learners~\cite{xie2020effect, CORBALAN2008733}, little research has explored how to balance control in the context of programming education, particularly from the perspectives of both learners and instructors.
In this paper, we refer to this shared control dynamic as \textbf{learner-system control}.

AI-supported learning systems can provide adaptive assistance in many dimensions, including: 
(1) the \textit{types of help} provided (e.g., hints, worked examples, and visualizations),
and (2) the \textit{level of help}, which refers to the amount of assistance provided (e.g., high-level subgoal labels versus a detailed breakdown of sub-subgoals).
Unlike the abstract \textbf{help features} described in RQ1, we use \textbf{types of help} to refer to the concrete ways of providing scaffolding.
Despite the growing number of AI-supported programming tools~\cite{le2013review}, 
little research has explored how to balance learner-system control across various types and levels of help.
To address this gap, we propose the second research question: 
\textbf{RQ2: What are novice learners' and instructors' preferences for \textit{learner-system control}?}

To answer these questions, we conducted a participatory design study with both undergraduate novice learners and programming instructors.
We created realistic programming practice scenarios for learners and instructional scenarios for instructors, asked them to design features for an ``ideal smart programming help'' tool to highlight the desired \textbf{help features}, and interviewed them regarding their preferred \textbf{learner-system control} mechanisms.
Based on the findings from the first two studies, we administered a survey on the abstract help features and learner-system control mechanisms and analyzed responses from 172 undergraduate students.
Finally, we present our findings and present design guidelines for AI-supported programming practice tools, especially for the learner-system control mechanisms.
\vspace{-10pt}
\vspace{-3pt}
\section{Related Work}
\vspace{-10pt}

\subsubsection{Learner-System Control Dynamics.}

A key challenge in adaptive learning is determining who controls the adaptation process: does the system make all decisions based on a student model, or does the learner control their learning experience~\cite{imhof2020implementation}? 
The former characterizes an \textit{adaptive} system, while the latter characterizes an \textit{adaptable} system~\cite{miller2005implications}. 
AI-supported learning systems can be placed on a spectrum between these extremes, depending on the balance of control between the system and its users~\cite{imhof2020implementation, nwana1990intelligent}.
A fully adaptive system can be problematic because its user model may misrepresent the learner~\cite{brusilovsky1998methods}, resulting in mismatches between system decisions and learner expectations and causing frustration~\cite{imhof2020implementation}. Conversely, fully adaptable systems can overwhelm learners at times, particularly novices, by providing too many choices~\cite{imhof2020implementation}.

The AIED community has studied ‘learner control,’ often with a focus on self-regulation and motivation~\cite{williams1993comprehensive}. 
Empirical studies have examined how learner control influences learning outcomes and behavior~\cite{CORBALAN2008733, xie2020effect, borchers2025learner}.
Corbalan et al.\cite{CORBALAN2008733} found that students who shared control with the system engaged more deeply in tasks than those in a system-controlled condition;
Xie et al.\cite{xie2020effect} demonstrated that granting students agency in a programming context increased engagement;
Recent research in intelligent tutoring systems (ITS)~\cite{borchers2025learner} indicates that learners highly value control as a means to enhance their learning experience and self-regulation.
While prior work offers valuable insights into control dynamics, these studies are often system-specific and have rarely investigated the \textit{preferences} of both learners and instructors. 
To fill this gap, our work is one of the first to directly study learners' and instructors' preferences using a design-focused approach, deriving guidelines for AIED systems in general.

Recent advances in AI, particularly in large language models, have enabled more personalized learning support~\cite{bernacki2021systematic, yan2024practical, abd2023large}. However, as Brusilovsky~\cite{brusilovsky2024ai} pointed out, despite growing interest in human-AI collaboration, ``\textit{the field of AIED is now lagging behind the work on user control in `big AI'}''. 
To emphasize the \textit{collaborative} decision-making process between learners and AI-supported systems, we refer to these evolving control dynamics as \textbf{learner-system control}.

\vspace{-13pt}

\subsubsection{Participatory Design in Educational Technology}
Participatory design (PD) comprises a set of human-centered methods that actively involve stakeholders in the design process~\cite{disalvo2017participatory}.
Engaging instructors leverages their domain expertise in teaching; however, involving learners is equally crucial
~\cite{konings2014participatory}.
PD methods have been used to create educational technologies in many fields, including math, history, and social skills~\cite{ismail2019participatory, borges2016participatory, pnevmatikos2020stakeholders, mack2019co, booker2016participatory}.
In computing education, rather than focusing on designing tools and systems for learners, PD methods have primarily been used to develop curricula for specific populations, such as to foster computational thinking skills for K-12 learners~\cite{sunday2024co} and to create culturally relevant curriculum~\cite{coenraad2022using}.
Other studies have directly integrated PD as learning activities for students~\cite{agbo2021co, limke2023empowering}. 
Our work is among the first to employ PD methods that directly engage both instructors and learners in the design process of AI-supported programming learning tools.

\vspace{-10pt}
\section{Participatory Design with Learners and Instructors}
\vspace{-10pt}

We conducted participatory design (PD) sessions with 15 undergraduate novice programmers and 10 instructors for both undergraduate and graduate programming courses.
All studies were conducted virtually via Zoom, received approval from the local institutional review board, and obtained participant consent.

\vspace{-12pt}
\subsubsection{PD with Learners}

We recruited 15 undergraduate participants who had either recently completed an introductory programming course or were enrolled to take it in the upcoming semester.
Prior to the PD sessions, participants filled out a demographic survey that included a standard self-efficacy survey for computing~\cite{wiggins2017you},
consisting of five questions rated on 7-point Likert scales (with 1 representing the lowest and 7 the highest).

\begin{figure}[h]
    \centering
    \vspace{-10pt}
    \begin{minipage}{0.38\textwidth}
        \includegraphics[width=\linewidth]{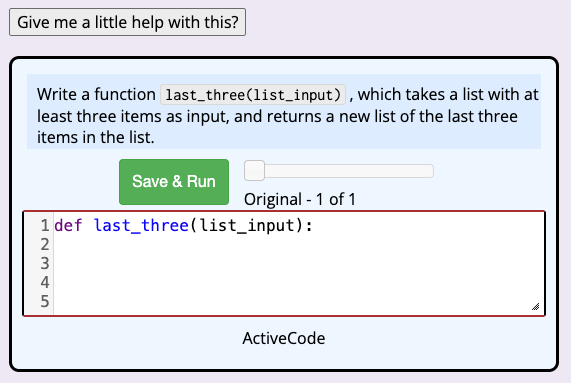}
        \subcaption{}
        \label{figsub:pd-interface}
    \end{minipage}
    \begin{minipage}{0.61\textwidth}
    \includegraphics[width=\linewidth]{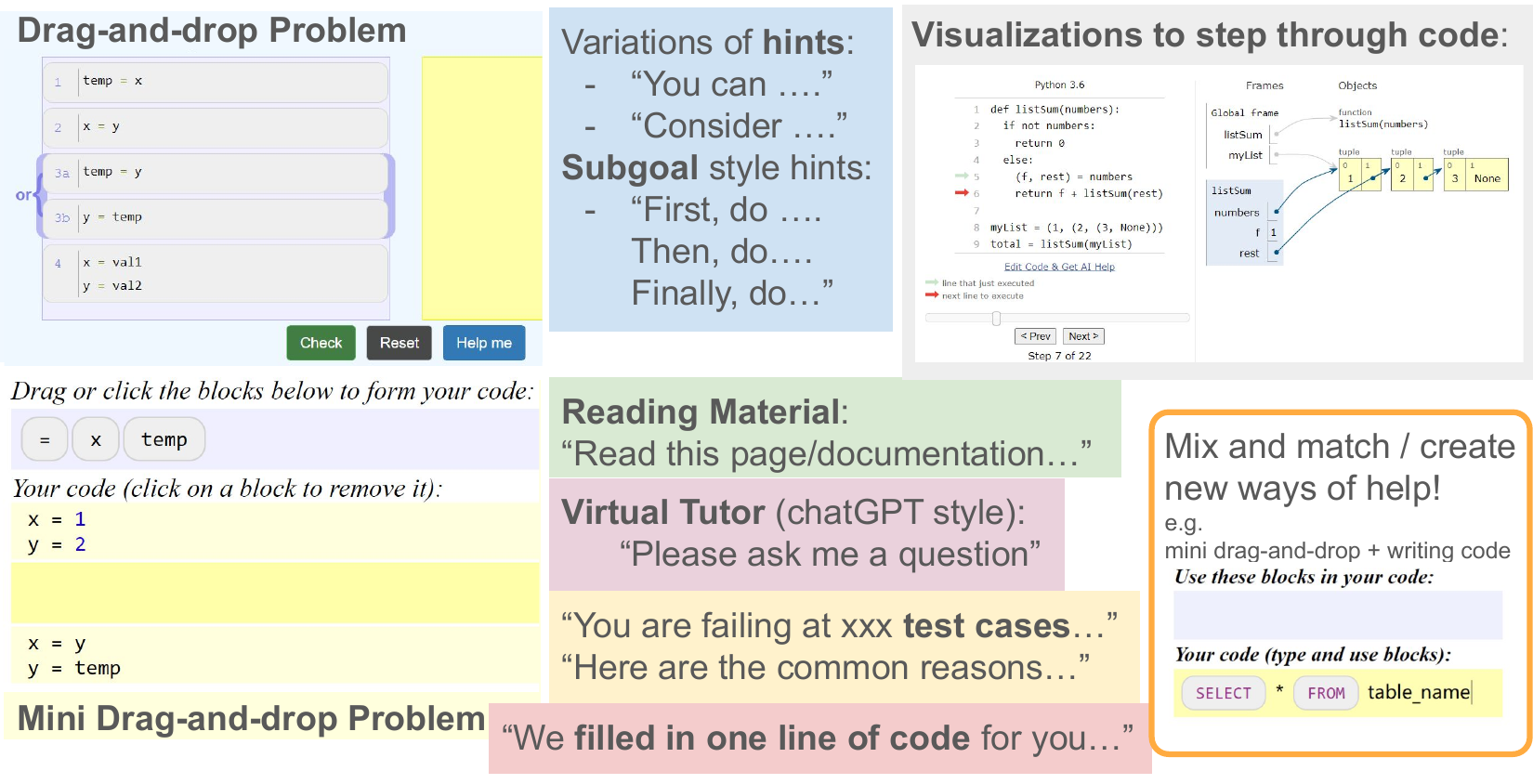}
        \subcaption{}
        \label{fig:slides}
    \end{minipage}%
\vspace{-7pt}
    \caption{(a): An example problem in the interface, with a help button on top. (b): Examples of different types of help, provided as inspirations for PD sessions.}
\vspace{-7pt}
\end{figure}

We established a context designed to elicit participants’ help-seeking behaviors.
First, learners were asked to attempt one or two introductory Python programming problems using a web interface (Fig.~\ref{figsub:pd-interface}) that allowed them to type and execute their code, view error messages from the Python interpreter, and receive unit test results.
The interface also featured a help button labeled ``Give me a little help with this?’' that was non-functional.
Participants were instructed to click the button when they desired assistance from the programming tool, which was designed to deliver the ``ideal'' help.
When they clicked the button, the researcher would: (1) inquire about the reason for the learner’s help request, (2) provide tutoring to assist in resolving the current issue, and (3) ask the learner to retrospectively describe the ``ideal'' help they would have preferred from the tool in the absence of a human tutor.
During the design brainstorming process, the researcher presented learners with common methods of providing help in existing programming practice tools (fig.~\ref{fig:slides}).

\vspace{-7pt}
\vspace{-7pt}

\vspace{-8pt}
Then, the researcher interviewed the participants for their desired \textbf{learner-system control} model for the programming tool.
The participants used an interface design tool to help articulate their desired design.
Learners were asked to consider two dimensions of the control model: (1) the fading of scaffolding as they become more experienced (\textit{levels of help}), and (2) managing their preferences for the different types of help available (\textit{types of help}).





\vspace{-12pt}
\subsubsection{PD with instructors}

We recruited 10 programming course instructors, including both faculty members and graduate student instructors.
We compiled learners’ help-seeking events from the PD sessions into a document (screenshots of learners’ code at the moment they pressed the help button).
For each event, instructors were asked to: (1) describe what the student was most likely to need help with, and (2) design the ideal support that the tool could provide.
Instructors were also provided with help examples (fig.~\ref{fig:slides}) as design inspirations.
Finally, the researcher posed the same interview questions regarding the instructors’ desired learner-system control model for the programming tool, focusing on both the \textit{types of help} and the \textit{levels of help}.

\vspace{-15pt}
\subsubsection{Data Analysis}

We collected screen and audio recordings for each session.
After anonymizing and transcribing the recordings, we conducted thematic analysis~\cite{terry2017thematic} on the transcripts, with the help of screen recordings to understand the context.
We developed one codebook for learners and another for instructors.
One researcher, responsible for all observation studies, performed an initial open coding to develop both codebooks.
Subsequently, two researchers independently coded a subset of the transcripts (two from learners and one from instructors) using the respective codebooks.
The researchers then discussed any discrepancies and refined the codebooks accordingly.
Finally, the two researchers independently coded two additional transcripts from both learners and instructors.
The two researchers reached a Krippendorf's alpha above the recommended inter-rater agreement threshold of $\alpha > 0.80$~\cite{krippendorff2018content} ($\alpha_{learner} = 0.82$, $\alpha_{instructor} = 0.81$).



\vspace{-10pt}
\section{Findings from Participatory Design Sessions}
\vspace{-6pt}


\subsection{Findings from Learners}
\vspace{-6pt}
Fifteen undergraduate participants (L01-L15) participated in the study.
Eight participants were information majors, six were computer science majors, and one was a business administration major.
Thirteen participants had just completed their sophomore year, and two had just completed their freshman year.
Nine participants identified as male, five as female, and one as non-binary.
The participants’ average self-reported self-efficacy was 5.03 (SD = 1.94) on a 7-point scale, where 1 indicates the lowest and 7 the highest level.

\vspace{-15pt}
\subsubsection{Help-Seeking Events.} 
On average, participants requested help 1.2 times per problem, and their reasons for seeking help were categorized into three codes.


\textit{When seeking an answer to a specific question.}
Eight (53\%) learners sought help when they had a specific question, such as requesting clarification on the problem or an explanation for a syntax error.
Learners did not hesitate long before pressing ``help'' in this case, knowing that they should receive the \textit{ideal} help from the tool in the study.
However, they worried that in a real setting, the tool might fail to accurately recognize their specific questions, potentially providing irrelevant or incorrect information.

\textit{When they attempted to implement a solution, encountered a problem, and were unable to resolve it despite believing they were on the right track.}
Nine (60\%) learners were categorized under this reason.
They attempted their ideas but encountered logical or syntactical problems that they failed to resolve, such as syntax errors or unexpected outputs.
Unlike the previous case, they were not sure what they needed to resolve the issue.
They generally delayed requesting help, because they wanted to continue making attempts: \myquote{at least I wanna try some of those (ways I know) before I ask for help.} (L02)
Eventually, when minor details became too frustrating, they asked for help.
They preferred not to receive a direct answer from the tool, so they could feel that their effort to resolve the problem paid off.



\textit{When unsure of how to begin and in need of guidance.}
Six (40\%) learners asked for help when they were uncertain about how to start the problem.
These learners were uncertain about the type of help required and had not yet attempted any solution.
Although some learners had preliminary ideas, they lacked confidence or recognized potential flaws in their planned solutions.
As a result, they decided to request help rather than proceed independently.
Learners wanted more help in this situation, expressing doubt about their ability to complete the problem, and wanted to make sure they were learning the desired way to code:
\myquote{Especially with getting started, I'd like to see how the textbook does it... it gives that sort of efficient logic, rather than sometimes my inefficient logic still works, but it's not always the best way.} (L10)

\vspace{-12pt}
\subsubsection{Preferences for Help Features.}
Five themes emerged from the PD sessions.
\begin{enumerate}
\vspace{-5pt}
    \item \textit{Provide feedback relevant to the student code} (n=15, 100\%). 
    Desired features include locating syntax or logic errors in their code, explaining failed test cases, and walking them through their code.
    \item \textit{Avoid providing the solution directly} (n=8, 53\%). 
    The learners did not want the tool to provide a solution directly: 
    \myquote{The help that I would be receiving would almost be meaningless... Because I'm not getting that conceptual understanding that I need.} (L13)
    \item \textit{Provide visual components beyond plain text} (n=4, 27\%).
    Learners referred to a variety of things as ``visual'', such as Parsons problems (i.e. rearranging mixed-up code blocks), debuggers, and code tracing tools. While some valued the graphics, others highlighted the need for visuals that effectively organize or chunck textual information.
    \item \textit{Be encouraging} (n=2, 13.4\%). 
    Learners noted that encouragement is not only for boosting motivation, but also helps prevent them from doubting themselves and starting over when their current code is on the right track.
    \item \textit{Provide information about and from peers} (n=2, 13.4\%). Desired features included showing common misconceptions, offering alternative solutions from peers, and retrieving relevant discussions on Q\&A platforms such as Piazza. 
\end{enumerate}
\textit{}

\vspace{-20pt}
\subsubsection{Preferences for learner-system control.}
\begin{figure}[h!]
    \centering
    \includegraphics[width=\linewidth]{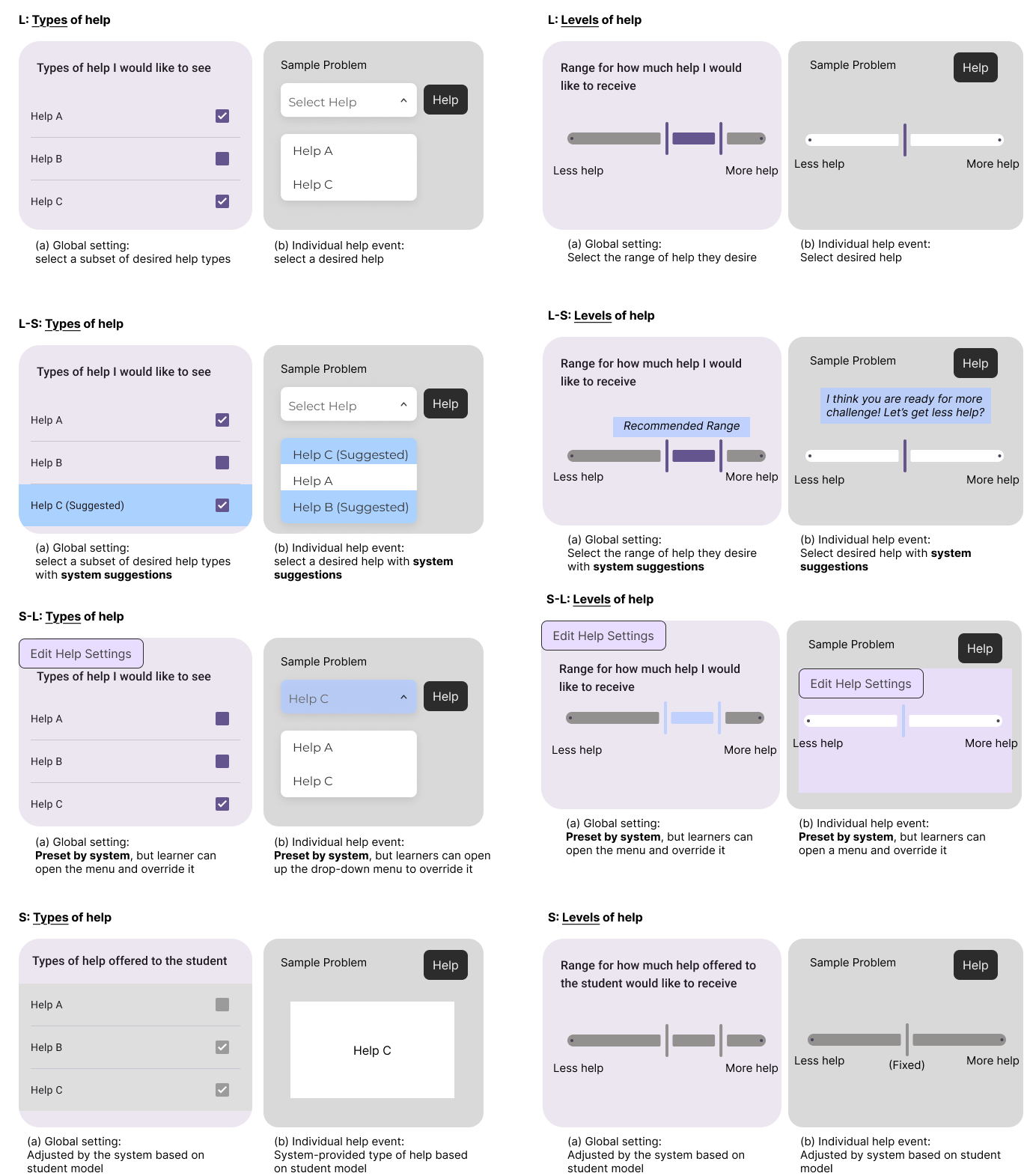}
    \caption{Example UIs that implement different learner-system control models.}
    \label{fig:learner-system-control}
\vspace{-20pt}
\end{figure}

We identified four distinct models of learner-system control.
Table~\ref{tab:learner-system-control} outlines each model and the implementation for \textit{types of help} and \textit{levels of help}, and fig.~\ref{fig:learner-system-control} provides example interface designs for implementing the control models.

\begin{table}[]
\vspace{-20pt}
\caption {Learner-system control models for \textit{types of help} and \textit{levels of help}} \label{tab:learner-system-control} 
\hspace{-15pt}
\begin{tabular}{p{1cm} p{3cm} p{0cm} p{4.5cm} p{0cm} p{4.5cm}}
\hline
Model & Explanation  && Controlling Types of Help && Controlling Levels of Help \\
\hline
\textbf{L} & \textbf{Learners have full control} and are not influenced by the AI system.  && The learner selects from a wide range of help types and receives only the type they choose, without system recommendations.     && The learner manually adjusts scaffolding, either increasing or decreasing the level of support as needed.                                                                                                                                 \\
\textbf{L-S} & \textbf{learners have major control}, but the \textit{AI system provides recommendations}.                       && The learner selects a preferred help type, while the system subtly suggests alternative types to assist decision-making.  && The learner manually adjusts scaffolding while the system provides gentle suggestions based on the student model.                     \\
\textbf{S-L} & The \textbf{AI system has major control}, but \textit{learners can override} system decisions. && The system recommends a help type based on learner progress, but the learner can override and choose a different type. && The system adjusts levels of scaffolding automatically, but learners can override these decisions when necessary. \\
\textbf{S} & The \textbf{AI system has full control}, and the learners receive help as determined by the system.     && The learner receives the one help type recommended by the system. No learner-initiated adjustment is needed or supported.  && The system fully controls scaffolding, automatically adjusting support based on learner progress, without learner overrides.                                                                                                                            \\ \hline
\end{tabular}
\vspace{-17pt}
\end{table}

In our study, learners were categorized based on their preferences regarding learner-system control.
\textbf{Three (20\%) learners preferred the L model}, and were confident about their ability to choose the appropriate type and level of help.
They had strong preferences towards the types and levels of help to receive, and wanted to manage their learning experience.
\textbf{Eight (53\%) learners fell into the L-S model}, wanting to be able to make the decisions, but would like to receive information provided by the system to assist with their decision.
For instance, participants L02 and L09 favored having predefined ‘go-to’ help types, yet still wanted system suggestions for individual help events.
\textbf{Two learners (13\%) preferred the S-L model} and would like the system to make decisions, 
but were concerned that the system would fail to provide the types of help that they wanted.
For example, L07 preferred having access to a menu offering access to all available help types, while L12 suggested a button to switch to an alternative help type if the provided one was unsatisfactory.
They were particularly concerned that the system would take away too much help when they needed it, therefore making the problem too difficult for them.
\textbf{Only one learner (6.7\%) preferred the S model.}
They felt that if the system is smart enough, they should not need to expend time and effort deciding on the type or level of help.

Beyond the control models, several additional considerations emerged during the PD sessions.
Some learners (n=2, 13\%) worried they might lack self-discipline at times, ending up abusing their power to control the help provided to them.
Additionally, some learners (n=4, 27\%) mentioned that they did not want to make decisions for every help-seeking event, suggesting that the frequency of learner-initiated control should be moderated.
Due to time constraints, we were unable to explore this issue in detail.

\vspace{-13pt}
\subsection{Findings from Instructors}
\vspace{-6pt}

\subsubsection{Desired help features.}
Although many instructors offered suggestions similar to those of the students, two unique recommendations emerged.

\begin{enumerate}
\vspace{-5pt}
    \item Promoting good practices. (n=5, 50\%). 
    Instructors recommended that the tool emphasize best programming practices, e.g. following good variable naming conventions and using print statements for debugging.
    \item Provide information on individual learning progress (n=2, 20\%). 
    Instructors proposed features that offer reminders of how learners previously resolved similar issues. They believe this could help learners feel more confident and encourage them to connect what they have learned in the past to what they are currently learning.
\vspace{-5pt}
\end{enumerate}

\vspace{-15pt}
\subsubsection{Learner-System control.}
The instructors’ interview responses were coded using the control model derived from the student interviews.
Most instructors (n = 6, 60\%) preferred S-L (mostly AI system control, but learners can override settings), and three (30\%) preferred L-S (mostly learner control, but the system provides recommendations).
Although instructors valued learner control, they expressed concern that requiring learners to make too many decisions might overwhelm them:
\myquote{It's important for students to have the ultimate control... [but] that's adding more stuff that they need to think about.} (I10)

Interestingly, one participant (I08) proposed a ``meta-control'' approach, customizing to students' preferences for how much control they would like to have:
\myquote{I will say that the most important thing is to understand what students want. So if the tool understands that the students want to make choices, then just give them the choices... If the students don't like making choices, it makes choices for the student.} (I08)

Additional considerations that are related to learner control were brought up by the instructors.
For example, all instructors (n=10, 100\%) mentioned that they would need input from students to provide the appropriate help, especially for determining the type of help, and would value the ability to \textbf{check in with students}.
Three (30\%) instructors expressed that they were \textbf{skeptical about the accuracy of student models} in AI-supported learning tools.
As a result, they believe that the learners should always have the final call on what help they receive, in case the provided system suggestion is incorrect.
Three (30\%) instructors mentioned their concerns that learners might ``game the system'' and abuse help, and suggested that the tool should be able to \textbf{adapt to learners' motivation in different contexts}.   
For example, I06 noted that for homework assignments, \myquote{most students will like the help that can help them get the full solution as fast as possible}, whereas when preparing for exams, they are more inclined to use sophisticated features to enhance learning.
Three (30\%) instructors also mentioned that they would like to be able to \textbf{directly control the settings} for the tool based on what they observe in class or receive insights on how learners use the tool.
Some wanted functionality that allows students to request help from instructors directly, such as shortcuts for students to send their current problem and incorrect solution to course communication channels like Slack and Piazza, or send them to the instructors directly during office hours.

\vspace{-10pt}
\section{Student survey}
\label{codesign-survey}
\vspace{-6pt}

\vspace{-6pt}
\subsection{Methods}
\vspace{-6pt}
We distributed an optional survey to learners in an introductory programming course and received 172 responses.
The survey consisted of questions for (1) self-efficacy, (2) perceived importance of help features identified from the PD sessions, and (3) preferred learner-system control models.
The survey began with a standard self-efficacy survey for computing~\cite{wiggins2017you} with five questions on 7-point Likert scales from 1 (lowest) to 7 (highest).
To reduce the length of survey, we extracted a total of five help features and created 5-point Likert scales for learners to rate the importance of each feature: (a)(Peer): Gives the learner information related to their peers, e.g., what other students’ solutions are and what problems other students are facing; (b)(Encouraging): Be encouraging, e.g. celebrating their progress on a problem even when not yet passing all test cases; (c) (Challenging): Challenges the learners, not giving them the answer directly; (d)(Visual): Has visual components aside from text, e.g. blocks or visualizations; (e)(Progress) Gives the learner information about their learning progress, e.g. similar problems they have solved or struggled with before. 

We asked learners to select their preferred learner-system control model in terms of (1) \textit{type of help} and (2) \textit{level of help} by describing the implementation of our proposed models in text.

\vspace{-15pt}
\subsection{Results}
\vspace{-5pt}
On a scale of 1 (low) to 7 (high), learners' average self-efficacy was 4.7 (SD=1.3).

\vspace{-12pt}
\subsubsection{Perceived importance of help features.}
\begin{figure}[h]
\vspace{-20pt}
    \centering
    \includegraphics[width=0.6\linewidth]{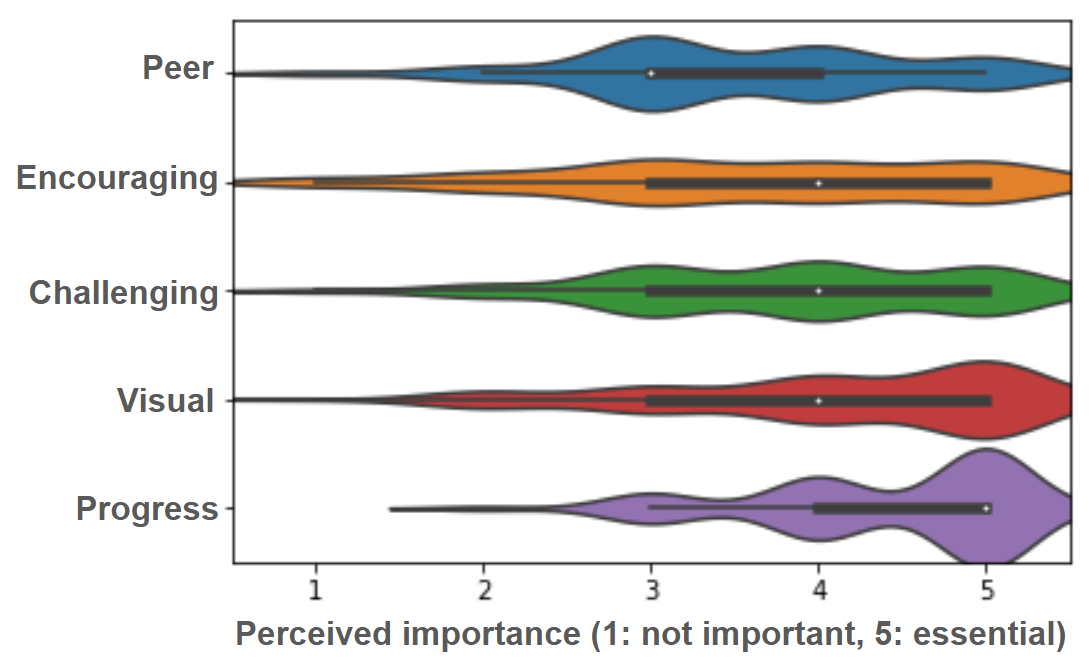}
    \vspace{-10pt}
    \caption{Violin plot of learners' perceived importance of the features.}
    \label{fig:survey-result-importance}
\vspace{-20pt}
\end{figure}

Learners rated the perceived importance (1 - not important, 5 -essential) of the abstract help features.
The average importance of providing peer information was 3.53 (SD=0.96), 
being encouraging was 3.56 (SD=1.18), 
being challenging was 3.78 (SD=0.99), 
having visual components was 4.09 (SD=1.03), 
and providing information about their learning progress was 4.38 (SD=0.79) (fig.~\ref{fig:survey-result-importance}).

We calculated the Pearson correlation between each learner's self-efficacy and their perceived importance of each of the five features.
Only the ``challenging the learner'' factor had a significant correlation with self-efficacy (p<.001), and the correlation was moderately positive~(r=0.36), which indicates that this was more important to those with higher computing self-efficacy.

\vspace{-15pt}
\subsubsection{Preferred learner-system control model.}
For the preferred learner-system control for the \textbf{type of help}, 39 (23\%) learners preferred L (learner-controlled), 92 (53\%) preferred L-S (mainly controlled by the learner), 34 (20\%) preferred S-L (mainly controlled by the AI system), and only 7 (4\%) chose S (AI-system-controlled).
The preferences for learner-system control models for the \textbf{level of help} were similar, where 38 (22\%) learners preferred the L model, 85 (49.4\%) preferred L-S, 43 (25\%) preferred S-L, and only 6 (3.5\%) chose S.
By assigning numerical values to these different learner-system control models (L: 0, L-S: 1, S-L: 2, S: 3), we performed a Pearson correlation between learners' preferred model and their self-efficacy.
At $\alpha = 0.01$ level, We found a significant correlation between preferences for learner-system control models for the \textbf{level of help} (p = 0.008), which was weak negative (r=-0.20).
This indicate that learners with higher self-efficacy desire more control for the \textbf{level of help} they receive from the AI system.
However, no statistically significant correlation was found between the preferred \textbf{type of help} with learners' self-efficacy.

\vspace{-10pt}
\section{Discussion}
\label{sec:discussion}
\vspace{-10pt}
In this section, we answer our research questions and offer design guidelines for designing help features and learner-system control mechanism for 
an ``ideal'' AI-supported learning tool, for programming learning and beyond.

\vspace{-13pt}
\subsubsection{Desired help features.}
\vspace{-6pt}
First, we provide design guidelines for \textbf{abstract features} that are desired from concrete help functionalities.

\begin{figure}[h]
    \vspace{-20pt}
    \centering
    \includegraphics[width=0.7\linewidth]{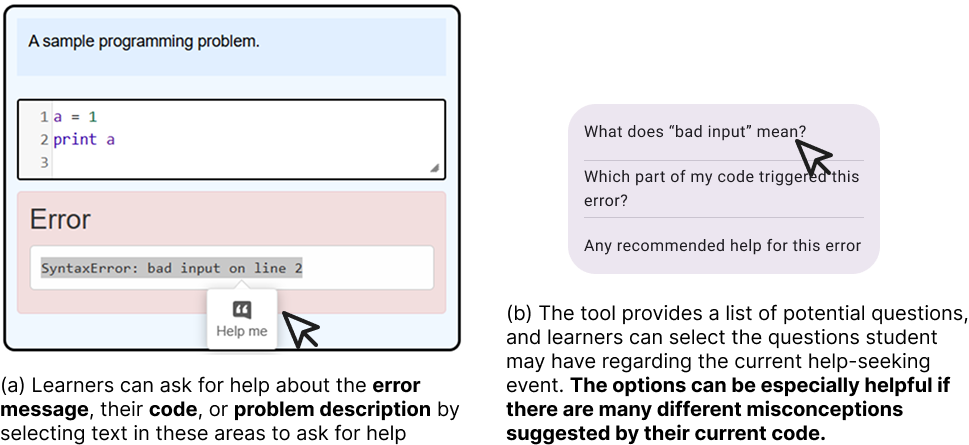}
    \vspace{-7pt}
    \caption{An example implementation of Guideline 1 with low interaction overhead.}
    \label{fig:sample-UI-explicit-input}
\end{figure}

\textbf{Guideline 1: Seek explicit input on why learners need help, and provide help based on learners' need. }
Our findings indicate that learners seek help from an AI-supported programming tool under three conditions: (a) when they have a specific question, (b) after unsuccessful attempts to resolve an issue, or (c) when they are unsure how to begin.
These help-seeking reasons reveal the type of information and assistance learners expect from the tool.
When a specific question is in mind, learners anticipate a direct and precise answer.
When confronting issues in their code, learners expected explanations that would guide them toward a resolution.
If a learner is unsure of how to start, they expect the tool to provide clear initial guidance.
However, even experienced instructors find it challenging to infer a learner’s intent only from their interactions with the programming interface. 
Fig.~\ref{fig:sample-UI-explicit-input} illustrates an example design that gathers explicit student input with minimal interaction overhead by (a) highlighting the area that is confusing for them or (b) asking learners to select an option that describes their current confusion.

\vspace{-18pt}

\textbf{Guideline 2: Provide personalized feedback on a micro level and a macro level.}
At the micro level, the tool should offer feedback specific to the current problem, such as annotating a learner’s solution to highlight errors or suggest improvements.
At the macro level, the tool should address a learner’s overall progress.
For instance, it can suggest similar problems they have encountered previously and remind them how they resolved a similar issue before to help them connect the dots.

\textbf{Guideline 3: Support diverse needs and provide space for customization.}
The perceived importance of many help features was very diverse.
While some participants repeatedly emphasized the value of an encouraging tool, a considerable amount of learners rated it as less important.
To accommodate these differing preferences and avoid frustrating or alienating users, AI-supported learning tools should offer customizable functionalities that adapt to individual needs.



\textbf{Guideline 4: Engage instructors by providing data access, control, and involvement opportunities.}
Many instructors also mentioned their desire to be more engaged with the system, instead of making the practice process an independent interaction between learners and the system.
The tool can provide analytical feedback to the instructors and allow instructors to customize the type of help available to learners. 
Additionally, the system could include shortcuts that allow learners to directly request assistance from instructors.

\vspace{-12pt}
\subsubsection{Learner-System Control}
We identified four preferred models of \textbf{learner-system control}~(see table~\ref{tab:learner-system-control} and Fig.~\ref{fig:learner-system-control}): full learner control (L), predominantly learner control with system suggestions (L-S),predominantly system control with learner override (S-L), and full system control (S).

\textbf{Guideline 5: Provide personalized learner-system dynamics.}
While the PD sessions suggested similar preferences for both the \textbf{type} and \textbf{level} of control, survey results revealed differences between them.
Learners with higher self-efficacy tended to desire greater control over the level of help provided, while this pattern was not observed with the types of help.
Although further investigation is warranted, it appears that preferences regarding the level of help may be more closely linked to cognitive factors such as self-efficacy and metacognitive skills, whereas preferences for the type of help may be driven more by individual taste.

\textbf{Guideline 6: Create context-aware learner-system control mechanism.}
Both learners and instructors voiced concerns about potential ‘gaming the system’ behavior. They also noted that the context in which programming practice occurs can significantly influence help-seeking behavior, especially for the \textbf{levels of help}.
For mandatory assignments where learners' main goal is to complete the problems, they are motivated to use help to reduce their time and effort.
For practicing their problem-solving skills and enhance their knowledge, they often hesitate to request help, thinking that receiving help might negatively affects their learning.
As a result, AI learning tools should be context-aware and adjust the range of control available for learners.

\vspace{-10pt}
\section{Limitations and Future Work}
\label{sec:limitations-future-work}

\vspace{-10pt}

One limitation is that we only studied self-reported data, which means that our work focused more on improving learners' satisfaction and engagement instead of learning outcome.
Learner participants in the PD sessions had a slightly higher average self-efficacy (mean=5.0) than the average responses from the survey (mean=4.7), indicating potential selection bias.
In terms of study design, although we provided examples in the participatory design sessions to help speed up learners' brainstorming process, their choices or preferences during the session could be affected by these examples.
While we provided a model to describe learner-system control dynamics, it is worth noting that there are factors not discussed in our proposed model, e.g. the desired frequency for learners to make control-related decisions, or the amount of information provided to learners regarding system decisions~\cite{xie2020effect}.
Future work can also explore more in using PD to create AI-supported learning tools.
Aside from having individual researcher-participant design sessions, future work can explore workshop activities that allow groups of learners to design culturally or community-relevant tools.
Larger workshops that engage learners and instructors at the same time can also provide an opportunity for different stakeholders to communicate their thoughts directly.
Future work can also expand outside of programming education to understand learners' and instructors' preferences in math or language learning.

\bibliographystyle{splncs04}
\bibliography{refs}
\end{document}